\documentclass[aps,pra,reprint,nofootinbib,groupedaddress,showpacs]{revtex4-1}
\usepackage{tabularx,amsmath,boxedminipage,graphicx,amssymb,bbm}
\usepackage{braket, dsfont}
\usepackage{setspace}
\usepackage{qcircuit}
\usepackage{titlesec}
\usepackage{color}
\usepackage{url}
\usepackage[final]{hyperref} 
\hypersetup{
	colorlinks=true,       
	linkcolor=blue,          
	citecolor=blue,        
	filecolor=magenta,      
	urlcolor=black         
}

\definecolor{Pr}{rgb}{0.4,0.3,0.9}

\begin{document}
\title{Continuous-variable gate decomposition for the Bose-Hubbard model}
\author{Timjan Kalajdzievski}
\author{Christian Weedbrook}
\author{Patrick Rebentrost}

\affiliation{Xanadu, 372 Richmond St W, Toronto, M5V 2L7, Canada}

\begin{abstract}
   In this work, we decompose the time-evolution of the Bose-Hubbard model into a sequence of logic gates that can be implemented on a continuous-variable photonic quantum computer. We examine the structure of the circuit that represents this time-evolution for one-dimensional and two-dimensional lattices. The elementary gates needed for the implementation are counted as a function of lattice size. We also include the contribution of the leading dipole interaction term which may be added to the Hamiltonian, and its corresponding circuit.
\end{abstract}

 \maketitle

\section{INTRODUCTION}
    
Quantum simulation of physical systems constitutes an important application for early quantum computing devices \cite{Feyn, Lloyd1, Lloyd2}. A quantum computer can be used
for the purpose of observing properties of that system which may be hard to obtain from direct experiments or classical computing. 
For example, such simulation may be used to determine the ground state energies of certain molecules or to simulate systems of molecules, which can be difficult to determine using a classical computer \cite{MolGrndState,Whitfield2011, HSimChem}.

Usually, the starting point is a reasonable model for the Hamiltonian of the physical system and mapping of that Hamiltonian into the degrees of freedom of the quantum simulator.  
Once a suitable mapping from the physical system has been found, the Hamiltonian time evolution operator is simulated by applying specific operations on the quantum device. 
The domain of Hamiltonian simulation examines the efficient implementation of Hamiltonians by considering their properties such as locality or sparsity. Often such simulation can be performed efficiently, that is polylogarithmically in the size of the Hilbert space and close to linear in the simulation time. For qubit quantum computers, such Hamiltonian simulations have been discussed in detail in \cite{HsimWalk, Berry2007, Berry2010, HSimChilds, Jordan2012, HSimChilds2, HSimSpectral, Childs2017, Haah2018}. 

The Bose-Hubbard model has been studied extensively, describing a system of bosonic particles trapped in an optical lattice \cite{dipole}. 
This model is simulated using various methods such as quantum Monte Carlo simulations \cite{BHSpace, BHSim1, BHSim2, BHSim3, BHSim4}. The purpose of most of these simulations has been to examine state transitions between a superfluid and a Mott insulator \cite{BHSim1, BHSim3, BHSim4, dipole, BHSim5}. The Bose-Hubbard model also has applications in examining the generation of entanglement \cite{entangle} and the creation of quantum magnetic insulators \cite{qmagnets}.  It has been shown that while the one-dimensional Bose-Hubbard may be easily simulated classically \cite{1d, BH1d}. However, the general problem of finding the ground state of a quantum system, including the Bose-Hubbard quantum system, is QMA-complete, and simulating the time evolution operator is BQP-complete when formalized as a decision problem \cite{qma1, qma2, qma3, Haah2018}. This means that there exists an efficient quantum algorithm that can accurately determine whether or not a given output was one produced from the Bose-Hubbard system, whereas it is believed that no such efficient classical algorithm exists. Here, efficient means that the algorithm scales as a polynomial in the size of the system.
    
A photonic continuous variable (CV) quantum computer utilizes the infinite-dimensional Hilbert space of the light field and can provide resource advantages compared to qubit quantum computers \cite{ChrisOverview}. Other advantages include room temperature computations and large-scale entanglement generation through the use of squeezers and low-cost components such as beam splitters and phase shifters \cite{Yokoyama2013}. Hamiltonian simulation can also be adapted for these continuous variable systems \cite{decompose}.
   
In this work, we discuss the simulation of the Bose-Hubbard Hamiltonian on a CV quantum computer.
For the Bose-Hubbard Hamiltonian, we show that a CV system allows for a straightforward mathematical decomposition into the required logic gates, as well as a circuit topology that allows for advantages in implementation.
We present the exact resource counts required to simulate the Bose-Hubbard Hamiltonian on a CV quantum computer. We consider the standard tunneling and on-site interaction terms of the Hamiltonian and also the addition of a dipole interaction term. We use Baker-Campbell-Hausdorff expansions in order to arrive at an elementary set of gates which are exponentials of powers of the position operator. This involves at most cubic and quartic single-mode gates. We also present the circuits that implement 1-D and 2-D Bose-Hubbard models of variable sizes. 

This paper is structured as follows. Section~\ref{Hamiltonian} presents the Hamiltonian and Sec.~\ref{Gate Decomposition} presents the standard relations for the decomposition of exponential polynomials of the position and momentum operators. In Sec.~\ref{Dipole Term}, we discuss the additional dipole term to the Hamiltonian. In Sec.~\ref{Circuit Implementations and Gate Counts}, we discuss the circuit implementations and the gate counts, as well as optical implementations of the gates and some potential sources of errors. In Sec.~\ref{Discussion} we offer a discussion and conclusion.
    
\section{BOSE-HUBBARD HAMILTONIAN}
\label{Hamiltonian}

\begin{figure}[h]
    \centering
    \includegraphics[width=0.45\textwidth]{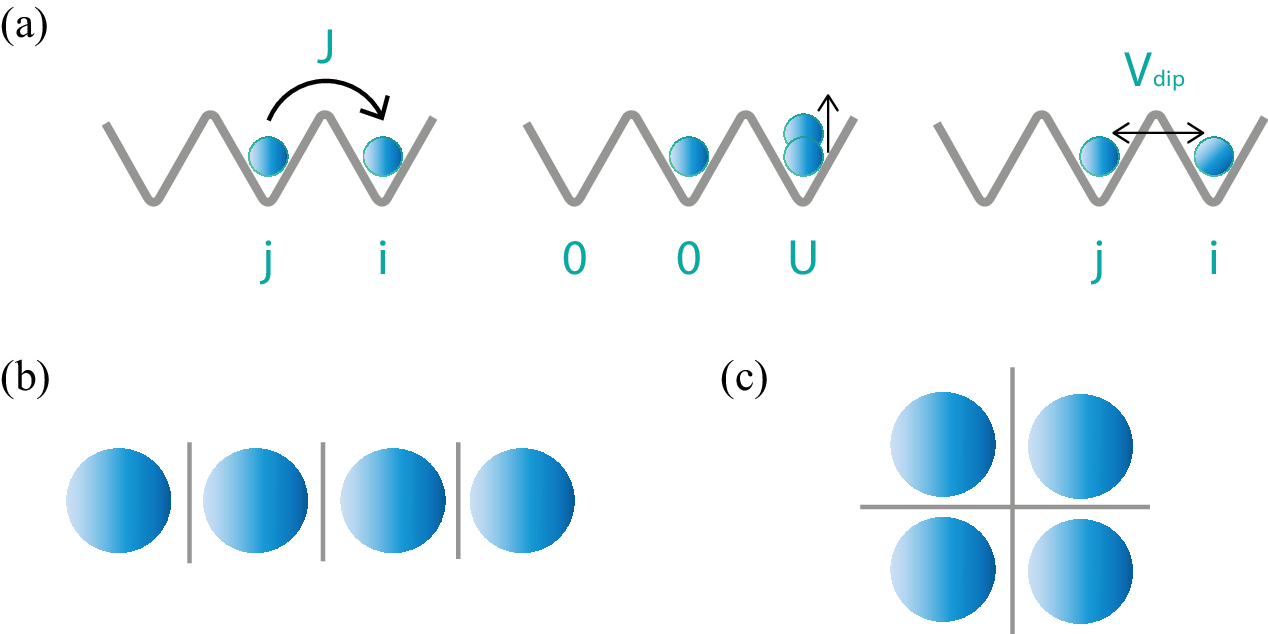}
    \caption{(a) A visualization of the effects of the terms in the Bose-Hubbard Hamiltonian. Here, $J$ is the tunneling coefficient which dictates the movement of particles from one site to a neighboring site, $U$ is the on-site interaction between two particles, and $V_{\rm{dip}}$ is the leading term of a dipole interaction between particles in neighboring sites. The part of the Hamiltonian which includes $V_{\rm{dip}}$ is discussed in more detail in Sec.~\ref{Dipole Term}. Also shown are two simple examples of lattices for which we examine the circuit implementations, (b) a one-dimensional four-node lattice and (c) a two-dimensional four-node lattice.}
    \label{BHFig}
\end{figure}

   The Bose-Hubbard Hamiltonian describes a system of bosonic particles trapped in an optical lattice of $N$ sites. Using notation from \cite{dipole}, it is given by
\begin{equation} \label{eqHamiltonian}
H = -\frac{J}{2}\sum_{\{i,j\}}\hat{a}^{\dagger}_{i}\hat{a}_{j} + \frac{U}{2}\sum_{i=1}^N\hat{n}_{i}(\hat{n}_{i} - 1),
\end{equation}
where the two terms with the factors $J$ and $U$ represent the tunneling of a particle in one site to a neighboring site, and the on-site interaction, respectively (see Fig.~\ref{BHFig} for a schematic). The bosonic creation (annihilation) operators are given by $\hat a^\dagger_i$ ($\hat a_i$) and the number operator is $\hat{n}_{i}=\hat{a}^{\dagger}_{i}\hat{a}_{i}$. The sum $\sum_{\{i,j\}}$ spans neighboring sites. Additional terms may be added to the Hamiltonian which come from dipole interactions \cite{dipole}. Methods to perform a gate decomposition for some of these terms are discussed in the appendix but first the terms in Eq.~(\ref{eqHamiltonian}) are examined in detail.
   The objective of this work is to find an appropriate implementation of quantum gates which can be used to simulate the evolution of this Hamiltonian $e^{itH}$ for a time $t$. In order to do this, $e^{itH}$ is decomposed into more elementary time evolution operators.

\section{GATE DECOMPOSITION} \label{Gate Decomposition}
    Note that the $J$ terms as well as part of the $U$ terms are of Gaussian order, therefore they may be efficiently implemented with linear optics. The non-Gaussian $U$ term may be further broken down into single-mode quadrature operations of quartic order, as well as Gaussian operations. This decomposition is now examined more precisely.
   First, the operators $\hat{a}^{\dagger}_{i}$, $\hat{a}_{i}$ and $\hat n_i$ are expanded in terms of position operators $\hat{x}_{i}$ and momentum operators $\hat{p}_{i}$ via
\begin{eqnarray}
\hat a_i &=& \hat x_i+i \hat p_i \nonumber ,\\
\hat a^\dagger_i &=& \hat x_i-i \hat p_i ,\\
\hat a^{\dagger}_i \hat a_i &=& \hat x_i^2 + \hat p_i^2+i[\hat{x}_{i}, \hat{p}_{i}] \nonumber .
\end{eqnarray} 
 In addition, the operators observe the commutator relation $[\hat{x}_{i}, \hat{p}_{i}] = i/2$.
 Considering these relations and neglecting a constant energy shift an expanded Hamiltonian is then written as
\begin{multline}
H = -J\sum_{\{i,j\}:i<j}\left(\hat{x}_{i}\hat{x}_{j} +  \hat{p}_{i}\hat{p}_{j} \right)+ \\
 \frac{U}{2}\sum_{i}\big(\left(\hat{x}^{4}_{i} + \hat{x}^{2}_{i}\hat{p}^{2}_{i} +\hat{p}^{2}_{i}\hat{x}^{2}_{i} + \hat{p}^{4}_{i} - \hat{x}^{2}_{i} - \hat{p}^{2}_{i} \right) + \\
 \left(- \hat{x}^{2}_{i} - \hat{p}^{2}_{i} \right) \big).
\end{multline}
We can simplify $\hat{x}^{2}_{i}\hat{p}^{2}_{i} +\hat{p}^{2}_{i}\hat{x}^{2}_{i}$ with a relation from \cite{decompose}
\begin{equation}
\hat{x}^{2}_{i}\hat{p}^{2}_{i} +\hat{p}^{2}_{i}\hat{x}^{2}_{i} = -\frac{4}{9}i[\hat{x}^{3}_{i},\hat{p}^{3}_{i}].
\end{equation}
As the time evolution to be simulated is $e^{itH}$, we can use the Lie product formula 
\cite{Childs2017} for sums of operators
$H=\sum_{j=1}^N H_j$,
\begin{equation}\label{eqLieProduct}
e^{i t \sum_{j=1}^N H_j} = \left( \prod_
{j=1}^N e^{i t H_j/K}  \right)^K + \mathcal R,
\end{equation}
where the choice of $K$ controls the size of the remainder $\mathcal R$ and thus gives the accuracy of the decomposition. The size of the remainder can be bounded by \cite{Childs2017}
\begin{equation} \label{eqErrorBound}
\Vert \mathcal R \Vert = O\left( \frac{N^2 t^2 \Lambda^2}{K} \right),
\end{equation} 
where $\Lambda:=\max_{j}\Vert H_j \Vert$ is the largest Hamiltonian norm. We discuss the choice of $K$ in Sec.~\ref{subsectionErrors} below. 
 In our case, we can write \begin{multline}
e^{itH} = \Bigg(\prod_{\{i,j\}:i<j} e^{-i\frac{t}{K}J\hat{x}_{i}\hat{x}_{j}}e^{-i\frac{t}{K}J\hat{p}_{i}\hat{p}_{j}} \\
\prod_{i} e^{i\frac{t}{K}\frac{U}{2}\hat{x}^{4}_{i}}e^{\frac{t}{K}\frac{2U}{9}[\hat{x}^{3}_{i},\hat{p}^{3}_{i}]}e^{i\frac{t}{K}\frac{U}{2}\hat{p}^{4}_{i}}
e^{-i\frac{t}{K}U\hat{x}^{2}_{i}}e^{-i\frac{t}{K}U\hat{p}^{2}_{i}} \Bigg)^{K} \\
 + \mathcal{R}.
\end{multline}  
The largest Hamiltonian norm here is at most $\Lambda = O({\rm poly} (J,U) )$, taken to be $O(1)$, as all terms involve the position and momentum operators \cite{decompose}.
 We can rotate every momentum operator into the position basis by a Fourier transform, denoted by $\mathcal{F}_{i}$ for mode $i$. For every polynomial $g$ we have
\begin{equation}
g(\hat{p}_i) = g(\mathcal{F}_{i}\hat{x}_i\mathcal{F}^{\dagger}_{i}) = \mathcal{F}_{i}g(\hat{x}_i)\mathcal{F}^{\dagger}_{i}.
\end{equation}
In addition, we can use commutator simulation via the relation \cite{decompose}
\begin{equation}\label{eqCommutatorSim}
e^{[A,B]\tau^{2}}=e^{iB\tau}e^{iA\tau}e^{-iB\tau}e^{-iA\tau}e^{iBt}e^{iA\tau}e^{-iB\tau}e^{-iA\tau} + O(\tau^4),
\end{equation}
to partition $e^{\frac{t}{K}\frac{2U}{9}[\hat{x}^{3}_{i},\hat{p}^{3}_{i}]}$ into terms involving 
$e^{i\left(\frac{t}{K}\frac{2U}{9}\right)^{1/2}\hat{x}^{3}_{i}}$
and $\mathcal{F}_{i}e^{i\left(\frac{t}{K}\frac{2U}{9}\right)^{1/2}\hat{x}^{3}_{i}}\mathcal{F}^{\dagger}_{i}$.
Note that in Eq.~(\ref{eqCommutatorSim}), $\tau$  is proportional to $(t/K)^{1/2}$, thus the error is proportional to $(t/K)^{2}$.
The final expanded form of the time-evolution operator is given by
\begin{multline}\label{eqTimeEvolution}
e^{itH} = \Bigg(\prod_{\{i,j\}:i<j} e^{-i\frac{t}{K}J\hat{x}_{i}\hat{x}_{j}}\mathcal{F}_{i}\mathcal{F}_{j}e^{-i\frac{t}{K}J\hat{x}_{i}\hat{x}_{j}}\mathcal{F}^{\dagger}_{j}\mathcal{F}^{\dagger}_{i} \\
\prod_{i} e^{i\frac{t}{K}\frac{U}{2}\hat{x}^{4}_{i}} \\
\mathcal{F}_{i}e^{i\left(\frac{t}{K}\frac{2U}{9}\right)^{1/2}\hat{x}^{3}_{i}}\mathcal{F}^{\dagger}_{i} e^{i\left(\frac{t}{K}\frac{2U}{9}\right)^{1/2}\hat{x}^{3}_{i}}\mathcal{F}_{i}e^{-i\left(\frac{t}{K}\frac{2U}{9}\right)^{1/2}\hat{x}^{3}_{i}} \\
\mathcal{F}^{\dagger}_{i} e^{-i\left(\frac{t}{K}\frac{2U}{9}\right)^{1/2}\hat{x}^{3}_{i}}\mathcal{F}_{i}e^{i\left(\frac{t}{K}\frac{2U}{9}\right)^{1/2}\hat{x}^{3}_{i}}\mathcal{F}^{\dagger}_{i} e^{i\left(\frac{t}{K}\frac{2U}{9}\right)^{1/2}\hat{x}^{3}_{i}}\\
\mathcal{F}_{i}e^{-i\left(\frac{t}{K}\frac{2U}{9}\right)^{1/2}\hat{x}^{3}_{i}}\mathcal{F}^{\dagger}_{i} e^{-i\left(\frac{t}{K}\frac{2U}{9}\right)^{1/2}\hat{x}^{3}_{i}} \\
\mathcal{F}_{i}e^{i\frac{t}{K}\frac{U}{2}\hat{x}^{4}_{i}}\mathcal{F}^{\dagger}_{i} 
e^{-i\frac{t}{K}U\hat{x}^{2}_{i}}\mathcal{F}_{i}e^{-i\frac{t}{K}U\hat{x}^{2}_{i}}\mathcal{F}^{\dagger}_{i} \Bigg)^{K} \\
 + O(\mathcal R).
\end{multline}
The error term that arises from Eq.~(\ref{eqCommutatorSim}) accumulates $K$ times, thus the contribution to the error in the final expression is proportional to $K\cdot \frac{t^2}{K^2} = \frac{t^2}{K}$ and can be absorbed into the existing error term.

\section{DIPOLE TERM} \label{Dipole Term}
   In case there is a dipole interaction between the bosons in the lattice an additional term may be added to the Hamiltonian \cite{dipole}, that is given by
\begin{equation}\label{eqHamiltonianDipole}
H_{nn} = V_{\rm{dip}} \sum_{\{i,j\}:i<j}\hat{n}_{i}\hat{n}_{j},
\end{equation}
   where this term is of leading order in dipole contributions, and corresponds to a nearest neighbor interaction. Other terms in dipole contributions are briefly examined in the appendix.
    
 Following the procedure from before, we can expand the Hamiltonian in terms of $\hat{p}$ and $\hat{x}$ operators, then rotate the $\hat{p}$s into $\hat{x}$s and decompose into single-mode quartic gates. Again to error $ O(N^2 t^2 /K)$, the sequence of gates includes the sequence of four Gaussian gates given by
\begin{equation}\label{eqDip1}
e^{-i\frac{t}{K}\frac{V_{\rm{dip}}}{2}\hat{x}^{2}_{i}}\mathcal{F}_{i}e^{-i\frac{t}{K}\frac{V_{\rm{dip}}}{2}\hat{x}^{2}_{i}}\mathcal{F}^{\dagger}_{i}e^{-i\frac{t}{K}\frac{V_{\rm{dip}}}{2}\hat{x}^{2}_{j}}\mathcal{F}_{j}e^{-i\frac{t}{K}\frac{V_{\rm{dip}}}{2}\hat{x}^{2}_{j}}\mathcal{F}^{\dagger}_{j},
\end{equation}
and four quartic terms given by
\begin{multline}\label{eqDip2}
e^{i\frac{t}{K}V_{\rm{dip}}\hat{x}^{2}_{i}\hat{x}^{2}_{j}}\mathcal{F}_{j}e^{i\frac{t}{K}V_{\rm{dip}}\hat{x}^{2}_{i}\hat{x}^{2}_{j}}\mathcal{F}^{\dagger}_{j}\mathcal{F}_{i}e^{i\frac{t}{K}V_{\rm{dip}}\hat{x}^{2}_{i}\hat{x}^{2}_{j}}\mathcal{F}^{\dagger}_{i}\\
\mathcal{F}_{i}\mathcal{F}_{j}e^{i\frac{t}{K}V_{\rm{dip}}\hat{x}^{2}_{i}\hat{x}^{2}_{j}}\mathcal{F}^{\dagger}_{j}\mathcal{F}^{\dagger}_{i}.
\end{multline}
Each of these two-mode quartic operators involving $\hat{x}^{2}_{i}\hat{x}^{2}_{j}$  can be decomposed into single-mode quartics and two-mode Gaussian operators using $12 \hat{x}^{2}_{i}\hat{x}^{2}_{j}=(\hat{x}_{i}-\hat{x}_{j})^4 + (\hat{x}_{i}+\hat{x}_{j})^4 - 2\hat{x}_{i}^4 -2\hat{x}_{j}^4$. In addition we  employ, $e^{i \hat p_i x_j} f(\hat x_i) e^{-i \hat p_i x_j} = f(\hat x_i + x_j) $ for appropriate functions $f(x)$.
This leads to the following relation
\begin{multline} \label{eqTwoModeQuartic}
e^{i\frac{t}{K}V_{\rm{dip}}\hat{x}^{2}_{i}\hat{x}^{2}_{j}} = \mathcal{F}_{i}e^{i2\hat{x}_{i}\hat{x}_{j}}\mathcal{F}^{\dagger}_{i}e^{i\frac{t}{K}\frac{V_{\rm{dip}}}{12}\hat{x}^{4}_{i}}\mathcal{F}_{i}e^{-i4\hat{x}_{i}\hat{x}_{j}}\mathcal{F}^{\dagger}_{i} \\
e^{i\frac{t}{K}\frac{V_{\rm{dip}}}{12}\hat{x}^{4}_{i}}\mathcal{F}_{i}e^{i2\hat{x}_{i}\hat{x}_{j}}\mathcal{F}^{\dagger}_{i}e^{-i\frac{t}{K}\frac{V_{\rm{dip}}}{6}\hat{x}^{4}_{i}}e^{-i\frac{t}{K}\frac{V_{\rm{dip}}}{6}\hat{x}^{4}_{j}}.
\end{multline}

\section{CIRCUIT IMPLEMENTATIONS AND GATE COUNTS} \label{Circuit Implementations and Gate Counts}
   In this section, we show the quantum circuits implementing the time evolution of the Bose-Hubbard model. We start by examining the circuit for a one-dimensional four-node lattice, and then examine the additional circuit of the dipole interaction term. We then generalize to two-dimensional lattices of size $n\times n$.
   
We first introduce several elementary operations. The following single-mode gates are used
\begin{eqnarray}
\mbox{
\Qcircuit @C=.5em @R=0em @!R {
& \gate{P(t)} & \qw &
\push{\rule{.3em}{0em}=\rule{.3em}{0em}} & \qw &\gate{e^{it\hat{x}^2}} &\qw
}}
\nonumber \\
\mbox{
\Qcircuit @C=.5em @R=0em @!R {
& \gate{V(t)} & \qw &
\push{\rule{.3em}{0em}=\rule{.3em}{0em}} & \qw &\gate{e^{it\hat{x}^3}} &\qw
}
}\\
\mbox{
\Qcircuit @C=.5em @R=0em @!R {
& \gate{Q(t)} & \qw &
\push{\rule{.3em}{0em}=\rule{.3em}{0em}} & \qw &\gate{e^{it\hat{x}^4}} &\qw
}
} \nonumber
\end{eqnarray}
where $P$ is a quadratic (shearing) gate consisting of the optical elements of squeezing and rotations; $V$ is the cubic phase gate; and $Q$ is the quartic gate.
The two-mode Cz, or C-PHASE, gate is given by
\begin{equation}
\mbox{
\Qcircuit @C=.7em @R=.5em @!R {
 &  \ctrl{2} & \qw &&&& \multigate{2}{e^{ig\hat{x}_1 \hat x_2}} & \qw &  \\
 & \rstick{g} &&& \push{\rule{.3em}{0em}=\rule{.3em}{0em}}&&&&& \\
& \ctrl{-2} & \qw &&&& \ghost{e^{ig\hat{x}_1 \hat x_2}} & \qw & 
}
}
\end{equation}
where we use a more generalized version of the Cz gate with a tunable (strength) parameter $g$. 

\begin{widetext}
\subsection{1-D Lattice Circuits}

   To present an example circuit for a single time step as in Eq.~(\ref{eqTimeEvolution}), we consider a 1-D lattice with four nodes as in Fig.~\ref{BHFig}(b). The circuit is given by 
\begin{equation}
\mbox{\Qcircuit @C=1em @R=.7em {
     & \multigate{1}{J} & \qw &  \qw & \qw & \qw & \gate{U} & \qw \\
     & \ghost{J} & \multigate{1}{J}  & \qw & \qw & \qw & \gate{U} & \qw\\
     & \qw & \ghost{J} & \multigate{1}{J} & \qw &\qw & \gate{U} & \qw \\
     & \qw & \qw & \ghost{J} & \qw & \qw & \gate{U} & \qw 
}}
\end{equation}
Here, the gate $J$ is given by
\begin{equation}\label{eqGateJ}
\mbox{
\Qcircuit @C=.7em @R=.5em @!R {
& \multigate{2}{J(g)} & \qw & & & \gate{\mathcal{F}^\dagger}  & \ctrl{2} &
\gate{\mathcal{F}} & \ctrl{2} & \qw \\
&&& \push{\rule{.3em}{0em}=\rule{.3em}{0em}} &&& \rstick{g} && \rstick{g}& \\
& \ghost{J(g)} & \qw & & & \gate{\mathcal{F}^\dagger}  & \ctrl{-2} & \gate{\mathcal{F}} & \ctrl{-2} & \qw
}
}
\end{equation}
The Cz gate is performed in between each pair of Fourier transform gates and $g$ is taken to be $g=tJ/K=:g_J$.
To simplify the $U$ gate we introduce a series of cubic and Fourier transform gates notated by $C$, given by the circuit
 \begin{equation}
\mbox{
\Qcircuit @C=.5em @R=0em @!R {
& \gate{C(t)} & \qw &
\push{\rule{.3em}{0em}=\rule{.3em}{0em}} & & \gate{V(t)} & \gate{\mathcal{F}^{\dagger}} & \gate{V(t)} & \gate{\mathcal{F}} & \qw
}
}
\end{equation}
The $U$ gate is then given by the circuit, with $g_U=\frac{tU}{K}$ and $g_C=(\frac{t}{K}\frac{2U}{9})^{1/2}$,
\begin{equation}
\mbox{
\Qcircuit @C=1em @R=.7em {
     & \gate{U(g_U,g_c)} & \qw &
\push{\rule{-.3em}{0em}=\rule{.5em}{0em}} & \gate{\mathcal{F}^{\dagger}} & \gate{P\left(g_U\right)} & \gate{\mathcal{F}} & \gate{P\left(g_U\right)} & \gate{\mathcal{F}^{\dagger}} & \gate{Q\left(\frac{g_U}{2 }\right)} & \gate{\mathcal{F}} & \gate{C(g_C)^4}& \gate{Q\left(\frac{g_U}{2 }\right)} & \qw
}
}
\end{equation}
The gates of each type needed for this circuit will be denoted
 in the form $(\mathcal{F}, P, V, Q, {\rm Cz})$. In the present case, we have $(\mathcal{F}, P(g_U), V(g_C), Q(g_U/2), {\rm Cz}(g))  = (60, 8, 32, 8, 6)$. Thus, for one time step, we need 60 Fourier gates, 8 quadratic gates (squeezers and rotations), 32 cubic gates, 8 quartic gates, and 6 Cz gates, with the given gate times $g, g_U,$ and $g_C$.

\subsection{Circuit For Dipole Term}
The additional dipole term may also be implemented in a circuit for a single time step in a 1-D lattice of $4$ nodes. This circuit is given by
\begin{equation}
\mbox{
\Qcircuit @C=1em @R=.7em {
     & \multigate{1}{V_{nn}} & \qw & \qw & \qw  \\
     & \ghost{V_{nn}} & \multigate{1}{V_{nn}}  & \qw & \qw \\
     & \qw & \ghost{V_{nn}} & \multigate{1}{V_{nn}} & \qw   \\
     & \qw & \qw  & \ghost{V_{nn}} & \qw 
}}
\end{equation}
To expand the $V_{nn}$ gate we introduce the decomposition of the two-mode quartic gate in Eq.~(\ref{eqTwoModeQuartic}) notated by $W$, which has the circuit
\begin{align}
&\Qcircuit @C=.5em @R=1.2em @!R {
& \multigate{2}{W} & \qw \\ 
&&& \push{\rule{.3em}{0em}=\rule{.3em}{0em}} \\
&\ghost{W} & \qw
} 
\! \! \! \! \! \! \! \! \! \! \! \! \! \! \! \! \! \! \! \! \! \! \! \! \! \! \! \! \! \! \! \! \! \! \! \! \! \!
&\Qcircuit @C=.5em @R=.5em @!R {
& \gate{Q\left(\frac{g_V}{3}\right)}  & \gate{\mathcal{F}^\dagger} & \ctrl{1} & \gate{\mathcal{F}} & \gate{Q\left(\frac{g_V}{6}\right)} & \gate{\mathcal{F}^\dagger} & \ctrl{1} & \gate{\mathcal{F}} & \gate{Q\left(\frac{g_V}{6}\right)} & \gate{\mathcal{F}^\dagger} & \ctrl{1} & \gate{\mathcal{F}} & \qw \\
& & & \rstick{2} &  &  &  & \rstick{-4} &  &  &  & \rstick{2} &  & 
\\
& \gate{Q\left(\frac{g_V}{3}\right)}  & \qw & \ctrl{-1} & \qw & \qw & \qw & \ctrl{-1} & \qw & \qw & \qw & \ctrl{-1} & \qw & \qw
}
\end{align}
Here, we take $g_V=tV_{\rm{dip}}/2K$. The $V_{nn}$ gate is then given by
\begin{equation}
\mbox{
\Qcircuit @C=1em @R=.7em {
     & \gate{P(g_V)} & \gate{\mathcal{F}^{\dagger}} & \gate{P(g_V)} & \gate{\mathcal{F}} & \multigate{1}{W} & \qw & \multigate{1}{W} & \qw & \gate{\mathcal{F}^{\dagger}} & \multigate{1}{W} & \gate{\mathcal{F}} & \gate{\mathcal{F}^{\dagger}} & \multigate{1}{W} & \gate{\mathcal{F}} & \qw \\
     & \gate{P(g_V)} & \gate{\mathcal{F}^{\dagger}} & \gate{P(g_V)} & \gate{\mathcal{F}} & \ghost{W} & \gate{\mathcal{F}^{\dagger}} & \ghost{W} & \gate{\mathcal{F}} & \qw & \ghost{W} & \qw & \gate{\mathcal{F}^{\dagger}} & \ghost{W} & \gate{\mathcal{F}} & \qw
}
}
\end{equation}

   Using a similar gate count notation we used previously, the dipole part of the circuit for the 1-D lattice will have a gate count of $(\mathcal{F}, P(g_V), Q\left(\frac{g_V}{3}\right), Q\left(\frac{g_V}{6}\right), {\rm Cz}(2), {\rm Cz}(-4)) = (108, 12, 24, 24, 24, 12)$. This means the total circuit including all of the $U$ and $J$ terms will have a gate count of
$(\mathcal{F}, P(g_U), P(g_V), V(g_C), Q\left(\frac{g_U}{2}\right), Q\left(\frac{g_V}{3}\right), Q\left(\frac{g_V}{6}\right), {\rm Cz}(g), {\rm Cz}(2), {\rm Cz}(-4))= (168, 8, 12, 32, 8, 24, 24, 6, 24, 12)$  for a single time step.
\subsection{2-D Lattice Circuits}
In this section, we discuss two-dimensional lattices of size $n \times n$. First, consider a $2\times2$ lattice with four total nodes as in Fig.~\ref{BHFig}(c). The circuit has the form 
\begin{equation}
\Qcircuit @C=1em @R=.7em {
     & \multigate{1}{J} & \gate{J} \qwx[2] & \qw & \qw & \gate{U} & \multigate{1}{V_{nn}} & \gate{V_{nn}} \qwx[2] & \qw & \qw & \qw \\
     & \ghost{J} & \qw & \gate{J} \qwx[2] & \qw & \gate{U} & \ghost{V_{nn}} & \qw & \gate{V_{nn}} \qwx[2] & \qw & \qw \\
     & \qw & \gate{} & \qw & \multigate{1}{J} & \gate{U} & \qw & \gate{} & \qw & \multigate{1}{V_{nn}} & \qw \\
     & \qw & \qw & \gate{} & \ghost{J} & \gate{U} & \qw & \qw & \gate{} & \ghost{V_{nn}} & \qw
}
\label{Lattice2x2}
\end{equation}
\end{widetext}
Here, we have introduced a new notation for the two-mode $J$ and $V_{nn}$ gates over two non-neighboring wires. This can be implemented on a circuit with only nearest neighbor coupling by swapping neighboring modes, applying the $J$ or $V_{nn}$ gates and then swapping back. For example,
\begin{equation}
\mbox{
\Qcircuit @C=.5em @R=0em @!R {
&&&& \lstick{1} & \gate{J} \qwx[2] & \qw & & & \lstick{1} & \qw & \multigate{1}{J} & \qw & \qw \\
&&&& \lstick{2} & \qw & \qw &
\push{\rule{.3em}{0em}=\rule{.3em}{0em}} & & \lstick{2} & \qswap & \ghost{J} & \qswap & \qw \\
&&&& \lstick{3} & \gate{} & \qw & & & \lstick{3} & \qswap \qwx & \qw & \qswap \qwx & \qw
}
}
\end{equation}
Note that the square box on the circuit indicates the other qumode that is being acted upon. This can similarly be done for an $n\times n$ lattice where, if the Bose-Hubbard model has nearest neighbor couplings, at most $n$ swaps are needed on either side of a gate. For an $n\times n$ lattice the first part of the circuit, which is the nearest neighbor pattern involving the $J$ gates, is given in Appendix \ref{appendixCircFull}.

We now show the final gate count for the $n\times n$ lattice. Following our notation as before, we also include a count for the number of swaps needed. For each $J$ gate the count is $(\mathcal{F},\rm{Cz}(g)) = (4, 2)$, and for the $n\times n$ lattice there are $2(n^{2}-n)$ $J$ gates and $2(n^{3}-n^{2})$ swaps, which gives us a gate count of $(\mathcal{F},\rm{Cz}(g), \text{SWAP}) = \left(8(n^{2}-n), 4(n^{2}-n), 2(n^{3}-n^{2})\right)$. As shown above, each $U$ gate has a count of 
$(\mathcal{F}, P(g_U), V(g_C), Q\left(\frac{g_U}{2 }\right))  = (12, 2, 8, 2)$
and in the lattice we have $n^{2}$ of them, giving a total count for the $U$ gates of 
$(\mathcal{F}, P(g_U), V(g_C), Q\left(\frac{g_U}{2 }\right))  = (12n^{2}, 2n^{2}, 8n^{2}, 2n^{2})$. Finally, each $V_{nn}$ gate has a count of $(\mathcal{F}, P(g_V), Q\left(\frac{g_V}{3}\right), Q\left(\frac{g_V}{6}\right), {\rm Cz}(2), {\rm Cz}(-4)) = (36, 4, 8, 8, 8, 4)$, and in the lattice the $V_{nn}$ gates follow the same pattern as the $J$ gates, so we have a total contribution from the $V_{nn}$ gates of $(\mathcal{F}, P(g_V), Q\left(\frac{g_V}{3}\right), Q\left(\frac{g_V}{6}\right), {\rm Cz}(2), {\rm Cz}(-4), \text{SWAP}) = (\big(72(n^{2}-n), 8(n^{2}-n), 16(n^{2}-n), 16(n^{2}-n), 16(n^{2}-n),$ $8(n^{2}-n), 2(n^3 - n^2)\big)$.
   
Therefore, the final gate count for our $n\times n$ lattice is 
\begin{multline}   
(\mathcal{F}, P(g_U), P(g_V), V(g_C), Q\left(\frac{g_U}{2}\right), Q\left(\frac{g_V}{3}\right), \\ Q\left(\frac{g_V}{6}\right), {\rm Cz}(g), {\rm Cz}(2), {\rm Cz}(-4), \text{SWAP}) = \\
 \big(92n^{2}-80n, 2n^2, 8(n^{2}-n), 8n^2, 2n^2, 16(n^{2}-n), \\ 16(n^{2}-n), 4(n^{2}-n), 16(n^{2}-n), 16(n^{2}-n), 4(n^{3}-n^{2})\big).
\end{multline}
Note that this is the gate count for each time step of length $t/K$ in the series of gates simulating $e^{iHt}$, as in Eq.~(\ref{eqTimeEvolution}) and Eqs.~(\ref{eqDip1}) to (\ref{eqTwoModeQuartic}).

\subsection{Optical Implementation}
   Note that the Gaussian elements of the circuits outlined in the previous sections can be implemented deterministically with linear optics whereas the higher-order gates are more complex and contain probabilistic elements. In this section, we briefly discuss the optical implementation of the various gates for completeness and note that the reader can find more information in the following citations.

 First, we examine the $J$ gate as in Eq.~(\ref{eqGateJ}). This circuit element consists of Fourier transforms and Cz gates which are single-mode Gaussian and multi-mode Gaussian operations and as such can be implemented with linear optics. Ref.~\cite{Gu2009} shows that any single-mode Gaussian operation can be implemented using rotations (or phase shifts), displacements, and either squeezing or shearing (squeezing and rotations). On the other hand, multi-mode Gaussian operations require the use of beam splitters. For the $J$ gate, the Fourier transforms are implemented simply with rotations of $\frac{\pi}{2}$, whereas the Cz gates require squeezers and the multi-mode transformation of beam splitters. More precisely, the Cz gate is given by \cite{linear}
 \begin{widetext}
\begin{equation}
\mbox{
\Qcircuit @C=.5em @R=.5em @!R {
& \ctrl{1} & \qw & & & \gate{S} & \multigate{2}{BS} & \qw \\
&&&\push{\rule{.3em}{0em}=\rule{.3em}{0em}} &&&&& \\
& \ctrl{-1} & \qw & & & \gate{S} & \ghost{BS} & \qw
}
}
\end{equation}
where squeezing is denoted by $S$ gates, and beam splitters by $BS$ gates. The tunable Cz gate will have similar components but needs the squeezing parameters and beam splitting ratios to be changed to fit our choice of $g_J=tJ/K$ \cite{GraphCalc}. Using the implementation for the tunable Cz gate along with the rotations that comprise the Fourier transforms, the $J$ gate can be optically implemented in the following way
\begin{equation}
\mbox{
\Qcircuit @C=.7em @R=.5em @!R {
& \multigate{2}{J(g)} & \qw &  & & \gate{R(-\frac{\pi}{2})}  & \gate{S(g)} & \multigate{2}{BS} &
\gate{R(\frac{\pi}{2})} & \gate{S(g)} & \multigate{2}{BS} & \qw \\
&&& \push{\rule{.3em}{0em}=\rule{.3em}{0em}} &&& &&& \\
& \ghost{J(g)} & \qw & & & \gate{R(-\frac{\pi}{2})}  &  \gate{S(g)} & \ghost{BS} & \gate{R(\frac{\pi}{2})} &  \gate{S(g)} & \ghost{BS} & \qw
}
}
\end{equation}
\end{widetext}

In order to implement higher-order gates we require more than the set of optical elements discussed above. The cubic phase operators denoted by the $V(t)$ gates in our circuits, are an example of these higher-order operations. To implement the cubic phase gate we add to our set of optical elements a photon counting measurement, which introduces the non-linearity needed. The full implementation then involves a displaced two-mode squeezed state for which $\hat{R}^\dagger \hat n \hat{R}$ (photon counting in a rotated basis) is measured on one arm. The desired cubic operation is then collapsed onto the second unmeasured mode \cite{Gottesman2001}. This is followed by a squeezing correction conditioned on the outcome of the photon number resolving detector followed by gate teleportation (all Gaussian elements). The initial two-mode squeezed state can be implemented with squeezing, beam splitters, and a phase shift which are all linear elements. 

For the optical implementation of our quartic gates ($Q$ gates in our circuits) we note that they may be expressed as a series of cubic gates by approximating in terms of commutators and using commutator simulation such as in Eq.~(\ref{eqCommutatorSim}) \cite{decompose, Marshall2015}. Thus, for quartic operations we do not need to add anything to our set of  linear optical components other than multiple photon counting. Note that in the case where a Kerr interaction is available, given by $e^{it\hat{n}_{i}^{2}}$, it may be used to directly implement the non-linear parts of the $U$ gates \cite{decompose, Braunstein2005, Brod2016}.

\subsection{A Note On Errors}
\label{subsectionErrors}
   When performing our gate decomposition and analyzing the makeup of our example circuits, note that all gate counts are given for a single Trotter time step. Let the desired accuracy of simulating  $e^{itH}$ be given by $\epsilon$. 
 The accuracy is dependent on the choice of number of time slices $K$, the total simulation time $t$, and the number of sites $N$.
   From Eq.~(\ref{eqErrorBound}), we can determine
    $K$ to achieve a given accuracy. 
    Such a $K$ is given by 
    \begin{equation}
    K= O \left (\frac{N^2 t^2}{\epsilon} \right ). 
    \end{equation}
The commutator simulation from Eq.~(\ref{eqCommutatorSim}) contributes at most in the same order as the sum formula Eq.~(\ref{eqErrorBound}). 
Our final product of operations for the Bose-Hubbard Hamiltonian is raised to the power of $K$, therefore we must repeat each circuit presented in this work $K$ times in order to get the desired error of $\epsilon$. 

   Another important source of error to discuss is the effect of finite squeezing. As discussed in the previous section, the optical implementation of the gates in our circuits will require the use of squeezing. In any experimental setup the squeezing will be finite and the end result with be dependent on a squeezing factor $s$ \cite{Menicucci2006, Gu2009}. For example, consider an optical implementation of the cubic phase gate where a photon counting measurement is made on a displaced two-mode squeezed state. To construct the two-mode squeezed state, two squeezed states, which ideally are zero-momentum eigenstates, are entangled. However, realistically the quadratures can only be finitely squeezed, in for example the momentum quadrature
\begin{equation}
\ket{0}_p \rightarrow \int dp\ e^{-(p)^{2}/(2s)} \ket{p}.
\end{equation}
The cubic phase gate is then modulated by a Gaussian envelope with zero mean and variance that depends on the squeezing factor $s$ \cite{Menicucci2006}. The result of this is a distortion effect which is inversely proportional to the amount of squeezing applied.

   In general, there are other experimental errors due to imperfections in the implementation. For example, actual photonic devices exhibit noisy state preparation, lossy interferometers and noisy and inefficient detectors. However, we leave such analysis for future work.

\section{DISCUSSION AND CONCLUSION} \label{Discussion}
   In this work, we have performed a decomposition of time-evolution under a bosonic Hamiltonian, namely the Bose-Hubbard Hamiltonian, into a set of elementary logic gates. Using our series of gates per-unit-time, we have presented a direct circuit implementation for a photonic quantum computer. The circuits discussed include a simple four-node, one-dimensional optical lattice for the Bose-Hubbard model and general two-dimensional lattices of size $n\times n$. Our final gate count is represented in terms of the number of gates of each type in our elementary set. For simulating the time-evolution of a Bose-Hubbard model to time $t$ and to error $\epsilon$ for an $n\times n$ lattice, the required number of gates is given by $(\mathcal{F}, P(g_U), P(g_V), V(g_C), Q\left(\frac{g_U}{2}\right), Q\left(\frac{g_V}{3}\right),$ $ Q\left(\frac{g_V}{6}\right), {\rm Cz}(g), {\rm Cz}(2), {\rm Cz}(-4), \text{SWAP}) = K \big(92n^{2}-80n, $ $2n^2, 8(n^{2}-n), 8n^2, 2n^2, 16(n^{2}-n), $ $16(n^{2}-n), 4(n^{2}-n), 16(n^{2}-n), 16(n^{2}-n), 4(n^{3}-n^{2})\big)$, where $K=O(n^2 t^2/\epsilon)$. The number of applications of each gate scales as a polynomial of the size of the lattice. 

    Even a small two-dimensional Bose-Hubbard model may be hard to simulate classically \cite{1d}. However, nearer-term photonic quantum computers may allow an implementation for a size $n$ where classical simulation is hard. The tuneable parameters $J$, $U$, and $V_{\rm{dip}}$ also allow a proof-of-principle experiment where we allow $V_{\rm{dip}}$ and $U$ to be much smaller than $J$. In the limit of infinitely small $U$ and $V_{\rm{dip}}$ terms, the circuit is fully Gaussian and implementable using only linear optics \cite{linear}, while at the same time being efficiently simulable classically. One can then systematically introduce non-linear gates to go beyond the classical simulable regime of the Bose-Hubbard model. This would allow photonic quantum computers with a limited number of non-linear gates to be used for the simulation of such a  physical system. It is also important to note that the dipole interaction term $H_{nn}$ used here is the leading term and more higher-order terms may be added \cite{dipole}. Implementing these higher-order terms in a CV system can be the subject of future work.

   Another interesting problem would be to find ways to decrease the final gate count. In the previous section, we used beam splitters to allow us to apply gates to two modes that are not neighbors in our circuit. Advances in photonic integrated circuit (PIC) design may remove the need for these beam splitters by using various topological techniques, such as crossing of photonic waveguides \cite{Nic2017}. Furthermore, the field of gate optimization in qubits is established but has yet to be established for CV systems. Therefore clever optimization tricks for CV systems for particular algorithms could also be constructed to reduce gate counts.

   The experimental implementation of higher dimensional gates is also potentially difficult and may require additional consideration. It is possible that it would be more useful to represent our quartic gates in terms of cubic gates and remove quartic gates from our elementary set \cite{decompose}. These higher dimensional gates may also require a feed forward implementation when decomposed in terms of lower dimensions \cite{nonlin}.

   It is also important to note that the procedure used in this work can be extended to other similar Hamiltonians. An efficiently simulable subclass of the Hamiltonian discussed here is the bosonic tight-binding Hamiltonian \cite{TBinding} with applications in condensed matter and solid state physics. The tight-binding Hamiltonian coupled to a bath of harmonic oscillators appears also in the study of exciton dynamics in photosynthetic complexes \cite{Exciton}. Simulating such systems can provide another application for continuous-variable photonic quantum processors.

\section*{ACKNOWLEDGMENTS}
   We would like to thank Pierre-Luc Dallaire-Demers, Tom Bromley, Zachary Vernon, and Nicolas Quesada for interesting discussions and helpful suggestions.

\bibliographystyle{apsrev}
\bibliography{BHD}

\onecolumngrid
\appendix

\section{J Gate Circuit For 2D Lattice of Arbitrary Size}
\label{appendixCircFull}

   The following circuit diagram is for the $J$ terms of the Bose-Hubbard Hamiltonian in Eq.~(\ref{eqHamiltonian}) applied to a two-dimensional, $n\times n$ lattice, cf. Sec. V.C. The dipole interaction term as in Eq.~(\ref{eqHamiltonianDipole}) will also have the same pattern, but will have gates notated with $V_{nn}$.

\vspace{5mm}

\Qcircuit @C=1em @R=.7em {
     &&&\lstick{1} & \multigate{1}{J} & \gate {J} \qwx[8] & \qw & \qw & \qw & \qw & \qw \\
     &&&\lstick{2} & \ghost{J} & \qw & \multigate{1}{J} & \gate{J} \qwx[8] & \qw & \qw & \qw &&\cdots \\ 
     &&&\lstick{3} & \qw & \qw & \ghost{J} & \qw & \multigate{1}{J} & \gate{J} \qwx[8] & \qw \\
     &&&\lstick{4} & \qw & \qw & \qw & \qw & \ghost{J} & \qw & \qw \\
     &&&&&& \cdot \\
     &&&&&& \cdot \\
     &&&&&& \cdot \\ \\
     &&&\lstick{n+1} & \qw & \gate{} & \qw & \qw & \qw & \qw & \qw \\
     &&&\lstick{n+2} & \qw & \qw & \qw & \gate{} & \qw & \qw & \qw &&\cdots&& \\
     &&&\lstick{n+3} & \qw & \qw & \qw & \qw & \qw & \gate{} & \qw \\
     &&&&&&& \cdot \\
     &&&&&&& \cdot \\
     &&&&&&& \cdot \\ 
}

\Qcircuit @C=1em @R=.7em {
     &&&&&&&&&&& \lstick{2n+1} & \multigate{1}{J} & \gate{J} \qwx[9] & \qw & \qw & \qw & \qw & \qw \\
     &&&&&\cdots&&&&&& \lstick{2n+2} & \ghost{J} & \qw & \multigate{1}{J} & \gate{J} \qwx[9] & \qw & \qw & \qw &&\cdots \\ 
     &&&&&&&&&&& \lstick{2n+3} & \qw & \qw & \ghost{J} & \qw & \multigate{1}{J} & \gate{J} \qwx[9] & \qw \\
     &&&&&&&&&&& \lstick{2n+4} & \qw & \qw & \qw & \qw & \ghost{J} & \qw & \qw \\ \\
     &&&&&&&&&&&&&& \cdot \\
     &&&&&&&&&&&&&& \cdot \\
     &&&&&&&&&&&&&& \cdot \\ \\
     &&&&&&&&&&& \lstick{3n+1} & \qw & \gate{} & \qw & \qw & \qw & \qw & \qw \\
     &&&&&\cdots&&&&&& \lstick{3n+2} & \qw & \qw & \qw & \gate{} & \qw & \qw & \qw &&\cdots \\ 
     &&&&&&&&&&& \lstick{3n+3} & \qw & \qw & \qw & \qw & \qw & \gate{} & \qw \\ 
}
\vspace{5mm}
\Qcircuit @C=1em @R=.7em {
     &&&&&&&&&&&\cdot&&&&&& \cdot \\
     &&&&&&&&&&&&\cdot&&&&& \cdot \\
     &&&&&&&&&&&&&\cdot&&&& \cdot \\ \\
     &&&&&&&&&&&&&&&& \gate{J} \qwx[8] & \qw & \qw & \qw & \qw & \qw & \rstick{n^{2}-n-2}  \\
     &&&&&&&&&&&&&\cdots&&& \qw & \qw & \gate{J} \qwx[8] & \qw & \qw & \qw & \rstick{n^{2}-n-1}  \\ 
     &&&&&&&&&&&&&&&& \qw & \qw & \qw & \qw & \gate{J} \qwx[8] & \qw & \rstick{n^{2}-n} \\ \\
     &&&&&&&&&&&&&&&&& \cdot \\
     &&&&&&&&&&&&&&&&& \cdot \\
     &&&&&&&&&&&&&&&&& \cdot \\ \\
     &&&&&&&&&&&&&&&& \gate{} & \multigate{1}{J} & \qw & \qw & \qw & \qw & \rstick{n^{2}-2} \\
     &&&&&&&&&&&&&\cdots&&& \qw & \ghost{J} & \gate{} & \multigate{1}{J} & \qw & \qw & \rstick{n^{2}-1} \\ 
     &&&&&&&&&&&&&&&& \qw & \qw & \qw & \ghost{J} & \gate{} & \qw & \rstick{n^{2}}
}


\section{Extended terms of Hamiltonian}

Using notation from \cite{dipole}, the occupation induced one-particle tunneling and the nearest-neighbor pair tunneling terms of the Bose-Hubbard Hamiltonian are respectively given by
\begin{equation} \label{eqHamiltonian}
H = -T\sum_{\{i,j\}}\hat{a}^{\dagger}_{i}(\hat{n}_{i} + \hat{n}_{j})\hat{a}_{j} + \frac{P}{2}\sum_{\{i,j\}}\hat{a}^{\dagger}_{i}\hat{a}^{\dagger}_{i}\hat{a}_{j}\hat{a}_{j}.
\end{equation}
Through the use of unitary conjugation with squeezing and displacement operations, the goal is to approximate the terms of the Hamiltonian by one or more gates for which a decomposition is known. This example illustrates a general method which can also be used to find a decomposition for other unitary operations. The method is inspired by techniques used in \cite{MabuchiKerr}.

Squeezing has the effect of multiplying $\hat{x}$ and $\hat{p}$ operators by a constant
\begin{align}
& S(\log\lambda)\hat{x}_{i}S^{\dagger}(\log\lambda) = \lambda \hat{x}_{i}, \\
& S(\log\lambda)\hat{p}_{i}S^{\dagger}(\log\lambda) = \lambda^{-1} \hat{p}_{i},
\end{align} 
and displacement has the effect of adding a constant (which will be labeled $\alpha$) \cite{Strawberry}. Applying both of these to the annihilation and creation operators maps them to a new effective operation
\begin{equation} 
\hat{a}^{\dagger} \rightarrow \hat{a}^{\dagger}_{\text{eff}} = \lambda \hat{x} - i\lambda^{-1}\hat{p} + \alpha.
\end{equation}
Applying this to the $P$ term of the Hamiltonian in Eq.~(\ref{eqHamiltonian}) first and then exponentiation in order to perform Hamiltonian simulation gives
\begin{equation} 
U_{P}^{\text{eff}} = e^{-i\tau H_{P}^{\text{eff}}} =  e^{-i\tau \frac{P}{2}\sum_{\{i,j\}}(\hat{a}^{\dagger}_{i, \text{eff}})^{2}(\hat{a}_{j, \text{eff}})^{2}}.
\end{equation}
The exponent can then be expanded in terms of $\hat{x}$ and $\hat{p}$, and simplified by grouping in terms of overall power of operators and constants $\lambda$ and $\alpha$
\begin{align}
 (\hat{a}^{\dagger}_{i, \text{eff}})^{2} (\hat{a}_{j, \text{eff}})^{2} &=   \\ \nonumber
& \; \; \left(\lambda^{2}\hat{x}_{i}^{2} - 2i\hat{x}_{i}\hat{p}_{i} + 2\lambda\alpha\hat{x}_{i} - \lambda^{-2}\hat{p}_{i}^{2} - 2i\alpha\lambda^{-1}\hat{p}_{i} + \alpha^{2}   \right)
\left(\lambda^{2}\hat{x}_{j}^{2} + 2i\hat{x}_{j}\hat{p}_{j} + 2\lambda\alpha\hat{x}_{j} - \lambda^{-2}\hat{p}_{j}^{2} + 2i\alpha\lambda^{-1}\hat{p}_{j} + \alpha^{2}   \right)  \\ \nonumber
 & = \big( \lambda^{4}\hat{x}_{i}^{2}\hat{x}_{j}^{2} + 2i\lambda^{2}\hat{x}_{i}^{2}\hat{x}_{j}\hat{p}_{j} - \hat{x}_{i}^{2}\hat{p}_{j}^{2} - 2i\lambda^{2}\hat{x}_{j}^{2}\hat{x}_{i}\hat{p}_{i} + 4\hat{x}_{i}\hat{p}_{i}\hat{x}_{j}\hat{p}_{j} + 2i\lambda^{-2}\hat{p}_{j}^{2}\hat{x}_{i}\hat{p}_{i} - \hat{p}_{i}^{2}\hat{x}_{j}^{2} - 2i\lambda^{-2}\hat{p}_{i}^{2}\hat{x}_{j}\hat{p}_{j}\\ \nonumber 
& \; \; + \lambda^{-4}\hat{p}_{i}^{2}\hat{p}_{j}^{2} + 2\alpha\lambda^{3}\hat{x}_{i}^{2}\hat{x}_{j} + 2i\alpha\lambda\hat{x}_{i}^{2}\hat{p}_{j} - 4i\alpha\lambda\hat{x}_{j}\hat{x}_{i}\hat{p}_{i} + 4\alpha\lambda^{-1}\hat{p}_{j}\hat{x}_{i}\hat{p}_{i} + 2\alpha\lambda^{3}\hat{x}_{i}\hat{x}_{j}^{3} + 4i\alpha\lambda\hat{x}_{i}\hat{x}_{j}\hat{p}_{j}\\ \nonumber 
& \; \; - 2\alpha\lambda^{-1}\hat{x}_{i}\hat{p}_{j}^{2}  - 2\alpha\lambda^{-1}\hat{x}_{j}\hat{p}_{i}^{2} - 2i\alpha\lambda^{-3}\hat{p}_{i}^{2}\hat{p}_{j} - 2i\alpha\lambda\hat{p}_{i}\hat{x}_{j}^{2} + 4\alpha\lambda^{-1}\hat{p}_{i}\hat{x}_{j}\hat{p}_{j} + 2i\alpha\lambda^{-3}\hat{p}_{i}\hat{p}_{j}^{2}+\alpha^{2}\lambda^{2}\hat{x}_{i}^{2} \\ \nonumber
& \; \; - 2i\alpha^{2}\hat{x}_{i}\hat{p}_{i} + 4\alpha^{2}\lambda^{2}\hat{x}_{i}\hat{x}_{j} + 4i\alpha^{2}\hat{x}_{i}\hat{p}_{j} - \lambda^{-2}\alpha^{2}\hat{p}_{i}^{2} - 4i\alpha^{2}\hat{p}_{i}\hat{x}_{j} + 4\alpha^{2}\lambda^{-2}\hat{p}_{i}\hat{p}_{j} + \alpha^{2}\lambda^{2}\hat{x}_{j}^{2} + 2i\alpha^{2}\hat{x}_{j}\hat{p}_{j} \\ \nonumber
& \; \; - \alpha^{2}\lambda^{-2}\hat{p}_{j}^{2} + 2\lambda\alpha^{3}\hat{x}_{i} - 2i\alpha^{3}\lambda^{-1}\hat{p}_{i} + 2\lambda\alpha^{3}\hat{x}_{j} + 2i\alpha^{3}\lambda^{-1}\hat{p}_{j} + \alpha^{4} \big).
\end{align}

Summing over all neighboring lattice sites implies that for a specific $i,j$ there will also be a term with swapped indices. Therefore, terms with alternate indices may be combined or canceled
\begin{align}
H_{P}^{\text{eff}} &= \frac{P}{2}\sum_{\{i,j\}}\big( \lambda^{4}\hat{x}_{i}^{2}\hat{x}_{j}^{2} - 2\hat{x}_{i}^{2}\hat{p}_{j}^{2} + 4\hat{x}_{i}\hat{p}_{i}\hat{x}_{j}\hat{p}_{j} + \lambda^{-4}\hat{p}_{i}^{2}\hat{p}_{j}^{2} + 4\alpha\lambda^{3}\hat{x}_{i}^{2}\hat{x}_{j} + 8 \alpha\lambda^{-1}\hat{p}_{j}\hat{x}_{i}\hat{p}_{i} - 4\alpha\lambda^{-1}\hat{x}_{i}\hat{p}_{j}^{2} \\ \nonumber
& \; \; + 2\alpha^{2}\lambda^{2}\hat{x}_{i}^{2} + 4\alpha^{2}\lambda^{2}\hat{x}_{i}\hat{x}_{j} + 4\alpha^{2}\lambda^{-2}\hat{p}_{i}\hat{p}_{j} - 2\alpha^{2}\lambda^{-2}\hat{p}_{i}^{2} + 4\alpha^{3}\lambda\hat{x}_{i} + \alpha^{4} \big) \\ \nonumber
 &= \frac{\lambda^{4}P}{2}\sum_{\{i,j\}}\big(\hat{x}_{i}^{2}\hat{x}_{j}^{2} +  4\alpha\lambda^{-1}\hat{x}_{i}^{2}\hat{x}_{j} + 2\alpha^{2}\lambda^{-2}\hat{x}_{i}^{2} + 4\alpha^{2}\lambda^{-2}\hat{x}_{i}\hat{x}_{j} + 4\alpha^{3}\lambda^{-3}\hat{x}_{i} + \mathcal{O}(\lambda^{-4}) \big),
\end{align} 
thus, to leading order $U_{P}^{\text{eff}} \approx  e^{-i\tau \frac{\lambda^{4}P}{2}\sum_{\{i,j\}}\hat{x}_{i}^{2}\hat{x}_{j}^{2}}$. The original unitary operator to be decomposed can be expressed in terms of the effective operator
\begin{equation}
U_{P} = S_{i}(\log\lambda)S_{j}(\log\lambda)D_{i}(\alpha)D_{j}(\alpha)U_{P}^{\text{eff}}D_{j}^{\dagger}(\alpha)D_{i}^{\dagger}(\alpha)S_{j}^{\dagger}(\log\lambda)S_{i}^{\dagger}(\log\lambda),
\end{equation}
where all of the operators on the right-hand side can now be decomposed to leading order in $\lambda$.

Note that the power of $\alpha$ increases when the overall order of the operators in each term decreases, and the power of $\lambda$ increases when there is a greater contribution of $\hat{x}$ operators over $\hat{p}$ operators. Choosing correctly $\alpha$ and $\lambda$ allows certain terms to have a much greater contribution than others \cite{MabuchiKerr}. For the $P$ term of the Hamiltonian, since the overall order is only four and fourth-order operators are not necessarily hard to decompose, the displacement operation can be left out to get a similar result:
\begin{align}
H_{P}^{\text{eff}} & = \frac{P}{2}\sum_{\{i,j\}}\big(\lambda^{4}\hat{x}_{i}^{2}\hat{x}_{j}^{2} + 4\hat{x}_{i}\hat{p}_{i}\hat{x}_{j}\hat{p}_{j} + \lambda^{-4}\hat{p}_{i}^{2}\hat{p}_{j}^{2} \big) = \frac{\lambda^{4}P}{2}\sum_{\{i,j\}}\big(\hat{x}_{i}^{2}\hat{x}_{j}^{2} +  \mathcal{O}(\lambda^{-4}) \big).
\end{align} 

To summarize the method, first map to effective operators by using squeezing and displacement (or just squeezing if overall order isn't an issue). Next, expand the expression in terms of quadrature operators $\hat{x}$ and $\hat{p}$, and simplify if the structure of the problem allows it. Then organize terms in order of power of $\lambda$, and as $\lambda \rightarrow \infty $ the leading order terms will have a much larger contribution. Choosing $\alpha$ to be a function of $\lambda$ can allow for lower order terms in powers of $\hat{x}$ and $\hat{p}$ to have a greater contribution as well. Finally, the original operator can be expressed as the effective operator with the opposite squeezing and displacement applied to it.

This process can also be used for the $T$ term of the Bose-Hubbard Hamiltonian in Eq.~(\ref{eqHamiltonian}). With just squeezing (and no displacement) the effective term is given by
\begin{align}
H_{T}^{\text{eff}} = -T\sum_{\{i,j\}}\big( \left(\lambda\hat{x}_{i} - i\lambda^{-1}\hat{p}_{i} \right)^{2} \left(\lambda\hat{x}_{i} + i\lambda^{-1}\hat{p}_{i} \right) \left(\lambda\hat{x}_{j} + i\lambda^{-1}\hat{p}_{j} \right)  \\ \nonumber 
+ \left(\lambda\hat{x}_{i} - i\lambda^{-1}\hat{p}_{i} \right)\left(\lambda\hat{x}_{j} - i\lambda^{-1}\hat{p}_{j} \right)\left(\lambda\hat{x}_{j} + i\lambda^{-1}\hat{p}_{j} \right)^{2}  \big).
\end{align}
After expanding and making use of similar symmetries that arose in the $P$ term, we can simplify to get 
\begin{align}
H_{T}^{\text{eff}} & = -T\sum_{\{i,j\}}\big( 2\lambda^{4}\hat{x}_{i}^{3}\hat{x}_{j} - 4\lambda^{2}\hat{x}_{j}\hat{x}_{i} - \hat{x}_{i}^{2}\hat{p}_{i}\hat{p}_{j} - \hat{p}_{i}^{2}\hat{x}_{i}\hat{x}_{j} - \hat{x}_{i}\hat{x}_{j}\hat{p}_{j}^{2} - \hat{p}_{i}\hat{p}_{j}\hat{x}_{j}^{2} \\ \nonumber
& + 2\hat{x}_{i}\hat{p}_{i}\hat{x}_{i}\hat{p}_{j} + 2\hat{x}_{i}\hat{p}_{i}\hat{x}_{j}\hat{p}_{i} + 2\hat{x}_{i}\hat{p}_{j}\hat{x}_{j}\hat{p}_{j} + 2\hat{x}_{j}\hat{p}_{i}\hat{x}_{j}\hat{p}_{j} -2\lambda^{-2}\hat{p}_{i}\hat{p}_{j} + 2\lambda^{-4}\hat{p}_{i}^{3}\hat{p}_{j} \big) \\ \nonumber \\ \nonumber
 & =  -T\lambda^{4}\sum_{\{i,j\}}\big(2\hat{x}_{i}^{3}\hat{x}_{j} - 4\lambda^{-2}\hat{x}_{j}\hat{x}_{i} + \mathcal{O}(\lambda^{-4}) \big).
\end{align}
Therefore, to leading order $U_{T}^{\text{eff}} \approx e^{i\tau T \lambda^{4}\sum_{\{i,j\}}2\hat{x}_{i}^{3}\hat{x}_{j}}$. Both $U_{P}^{\text{eff}}$ and $U_{T}^{\text{eff}}$ can be decomposed using techniques from \cite{ExactDecomp}.

\end{document}